\documentclass[12pt]{article}
\input epsf.sty
\def\be{\begin{equation}}
\def\ee{\end{equation}}
\def\ba{\begin{align}}
\def\ea{\end{align}}

\def\p{\partial}

\def\ops[#1]{\p_{#1} e^{-2\phi}}

\def\eq[#1]{equation (\ref {eq:#1})}
\def\Eq[#1]{Equation (\ref {eq:#1})}
\def\e[#1]{\ref {eq:#1}}
\def\at[#1]{| _{#1}}

\let\oldpercent\%\renewcommand{\%}{\scalebox{0.85}{\oldpercent}}

\begin{document}

\baselineskip=18pt

\begin{center}
{\Large \bf{To $B$ or not to be in single-trace $T\bar T$ holography}}

\vspace{10mm}

\textit{Amit Giveon and Daniel Vainshtein}
\break

Racah Institute of Physics, The Hebrew University \\
Jerusalem, 91904, Israel

\vspace{6mm}

giveon@mail.huji.ac.il\\
daniel.vainshtein@mail.huji.ac.il

\end{center}


\vspace{10mm}

\begin{abstract}

Superstring theory on black-strings backgrounds, corresponding to deformed, rotating BTZ black holes, formed
by $p$ fundamental strings or {\it negative} strings, is inspected.
For {\it non}-rotating black strings, in the positive case, it was shown in~\cite{Chakraborty:2024ugc}
that the excitation energy of a probe long string, plus its contribution to the black-string background,
is in harmony with single-trace $T\bar T$ holography, if the $B$-field vanishes at the origin.
Here, we investigate the superstring on deformed, {\it rotating} BTZ black holes, for both the positive and negative cases,
and find that for a particular value of the $B$-field, the harmony with single-trace $T\bar T$ holography persists.


\end{abstract}
\vspace{10mm}

\section{Introduction}

Single-trace $T\bar T$ holography is obtained from the $AdS_3/CFT_2$ one,
in perturbative superstring theory on $AdS_3$,
via a solvable irrelevant deformation,~\cite{Giveon:2017nie,Giveon:2017myj}.
On the $SL(2,R)$ worldsheet CFT, it is obtained by a truly marginal deformation,
which in the hologram amounts to deforming the $CFT_2$ with a certain $(2,2)$ quasi-primary operator.
The theory is referred to as single-trace $T\bar T$ holography,
since it shares many properties with $T\bar T$ deformed $CFT_2$,~\cite{Smirnov:2016lqw,Cavaglia:2016oda}.

In particular, the long-strings spectrum, which in the undeformed superstring on $AdS_3$~\footnote{Or
on a massless BTZ worldsheet CFT (if the boundary of $AdS_3$ is compactified on a circle).}
is identical to that of a symmetric orbifold,~\cite{Argurio:2000tb,Giveon:2005mi},
turns upon the deformation to that of a $p$-fold symmetric product, ${\cal M}^p/S_p$, of a  $T\bar T$ deformed $CFT_2$ seed,~\cite{Giveon:2017myj}.

In~\cite{Giveon:2017nie}, it was also claimed that black-strings solutions in the deformed massless BTZ theory,
correspond to high-energy states in the Ramond-Ramond sector of the hologram,
whose energy and momentum are equally split among the $p$ blocks of ${\cal M}^p/S_p$.
Later,~\cite{Apolo:2019zai,Chakraborty:2020swe,Chang:2023kkq,Chakraborty:2023mzc,Chakraborty:2023zdd},
this was shown in detail for general rotating black-strings geometries,
obtained by deforming BTZ black holes with mass and angular momentum, $M_{BTZ}$ and $J_{BTZ}$,
with a dimensionless deformation parameter, that we'll denote by $\lambda$,
both for the positive direction, $\lambda>0$, and for negatively deformed BTZ, $\lambda<0$.
In the hologram, it amounts to a single-trace $T\bar T$ deformation, with a dimensionless coupling~$\lambda$.~\footnote{The same
was claimed to be the property for deformed global $AdS_3$ geometry, which is dual to the NS-NS ground state of the
single-trace $T\bar T$ hologram,~\cite{Apolo:2019zai,Chang:2023kkq,Chakraborty:2023mzc}.}

Moreover, in~\cite{Chakraborty:2024ugc}, it was shown that for {\it non}-rotating black strings, $J_{BTZ}=0$,
in the positive coupling case, $\lambda>0$,
the excitation energy of a probe long string, plus its contribution to the black-string background,
is in harmony with single-trace $T\bar T$ holography, if the $B$-field vanishes at the origin.~\footnote{The same
was shown to be the property for probe long strings in deformed global $AdS_3$,~\cite{Chakraborty:2024ugc,Chakraborty:2024mls}.}

For more details on the present status of single-trace $T\bar T$ holography,
we refer the reader to~\cite{Chakraborty:2023mzc,Chakraborty:2023zdd,Chakraborty:2024ugc}, and references therein.

In this note, we inspect the spectrum of excited long strings in superstring theory on {\it rotating} black strings,
formed by $\lambda$-deforming BTZ black holes (with $J_{BTZ}>0$), for both $\lambda>0$ and $\lambda<0$.
The results for the positive and negative coupling cases are presented in sections 2 and 3, in turn.
In both cases, we find that the long-strings pattern of the spectrum is in agreement with that in an ${\cal M}^p/S_p$ theory,
if and only if the value of the $B$-field at the origin is appropriately fixed.~\footnote{The
bottom line is in~(\ref{bbbzero})--(\ref{btt}).}

Concretely, in superstring theory with the $B$ to be in single-trace $T\bar T$ holography,
the total energy of a long string,
consisting of its excitation energy plus its contribution
to the deformed BTZ backgrounds energy,
is deformed according to the $T\bar T$ formula of~\cite{Smirnov:2016lqw,Cavaglia:2016oda},
in a  block of a symmetric product,
for both positive and negative couplings,
and for any $J_{BTZ}$.

To emphasize the importance of the $B$-field in single-trace $T\bar T$ holography,
other choices in the literature are presented in section 4.
For those, the long-strings spectrum necessarily differ from that in single-trace $T\bar T$ holography,
and it ain't clear to us if such choices are possible in consistent perturbative string theory.

In an appendix, we present some details of calculations and arguments, which are used to obtain our results.

\section{Positive coupling}

Consider supersting theory on the deformed BTZ black holes background,
formed in the near $k$ $NS5$ branes on $S^1\times T^4$
with $p$ fundamental strings ($F1$) wrapping $S^1$, whose radius we denote by $R$, and let the strings carry momentum number
\be\label{nfone}
n_{F1}=J_{BTZ}\in Z_+~,
\ee
on this circle, where $J_{BTZ}$ is the angular momentum of the (undeformed) rotating BTZ black hole.~\footnote{We take
$J_{BTZ}\geq 0$, without loss of generality; in the BTZ limit it is equal to the momentum charge of the fundamental strings
that form it,~(\ref{nfone}), and it is fixed along the deformation.}
The deformed BTZ background appears in e.g. equations (2.17) with (2.18) of~\cite{Chakraborty:2023zdd}:~\footnote{In~\cite{Chakraborty:2023zdd},
$n_{F1}$ was denoted by $n$; here we keep $n$ for the momentum of an excited string below.}
\be\label{defbtz}
ds^2=-{N^2\over 1+{\rho^2\over R^2}}d\tau^2+{d\rho^2\over N^2}+{\rho^2\over 1+{\rho^2\over R^2}}(d\theta-N_\theta d\tau)^2~,\qquad \theta\simeq\theta+2\pi~,
\ee
\be\label{btautheta}
B_{\tau\theta}=B_{\tau\theta}^0
+{\rho^2\over r_5}\sqrt{\left(1+{\rho_-^2\over R^2}\right)\left(1+{\rho_+^2\over R^2}\right)}{1\over 1+{\rho^2\over R^2}}~,
\qquad B_{\tau\theta}^0\equiv B_{\tau\theta}(\rho=0)~,
\ee
\be\label{dilaton}
e^{2\Phi}={kv\over p}\sqrt{\left(1+{\rho_-^2\over R^2}\right)\left(1+{\rho_+^2\over R^2}\right)}{1\over 1+{\rho^2\over R^2}}~,
\qquad v\equiv {\rm Volume}(T^4)/\left(2\pi\sqrt{\alpha'}\right)^4~,
\ee
with~\footnote{In~\cite{Chakraborty:2023zdd},
the angular direction $\theta$ of the background~(\ref{defbtz}),(\ref{btautheta}) was denoted by $\varphi$;
below, $\varphi,\bar\varphi$ will denote worldsheet fields associated with the superconformal ghosts, instead.}
\be\label{withn}
N^2={\left(\rho^2-\rho_+^2\right)\left(\rho^2-\rho_-^2\right)\over r_5^2\rho^2}~, \qquad N_\theta={\rho_+\rho_-\over r_5\rho^2}~,
\ee
and
\be\label{k}
r_5\equiv\sqrt{\alpha' k}~.
\ee
This background describes a rotating black string in $2+1$ dimensions.~\footnote{See e.g.~\cite{Chakraborty:2023zdd} for its properties.}


A couple of comments are in order:

\begin{itemize}

\item
From the point of view of supergravity, $B_{\tau\theta}^0$ in~(\ref{btautheta}) can take any value.~\footnote{Since
it does not affect the value of $H=dB$.}
On the other hand, from the point of view of the worldsheet CFT, a.k.a. in perturbative string theory, generically,
the situation is not clear.~\footnote{In~\cite{Chakraborty:2023zdd}, $B_{\tau\theta}^0$ was chosen to be zero for any $\rho_\pm$;
below, we'll prefer a different choice, when $\rho_-\neq 0$.}

\item
The value of $B_{\tau\theta}^0$ will play a central role in this work.
We will argue below that it should satisfy
\be\label{restrict}
B_{\tau\theta}^0(R/\sqrt{\alpha'}\to\infty)\to 0~,
\ee
for consistency with the BTZ limit.

\item
The results in this paper apply to the superstring on deformed BTZ times any internal space ${\cal N}$.
The properties of ${\cal N}$ will not play any role below.~\footnote{In the concrete example above, ${\cal N}=SU(2)_k\times T^4$.}

\end{itemize}

Next, define,
\be\label{tx}
t\equiv{R\tau\over r_5}~,\qquad x\equiv R\theta~,\qquad{\phi\over r_5}\equiv\log\left(\rho\over R\right)~,
\ee
and
\be\label{bzero}
b_0\equiv B_{tx}(\rho=0)={r_5\over R^2}B_{\tau\theta}^0~,
\ee
as well as
\be\label{binfty}
b\equiv B_{tx}(\rho\to\infty)={r_5\over R^2}B_{\tau\theta}(\rho\to\infty)~.
\ee
Asymptotically, a.k.a. at $\rho/R\to\infty$, the black-string background~(\ref{defbtz})--(\ref{withn})  takes the form
\be\label{ds}
ds^2\to -dt^2+dx^2+d\phi^2~,
\ee
\be\label{b}
B_{tx}\to b=b_0+\sqrt{\left(1+{\rho_-^2\over R^2}\right)\left(1+{\rho_+^2\over R^2}\right)}~,
\ee
\be\label{phiphi}
\Phi\to -{\phi\over r_5}+const~,
\ee
a.k.a. it is time times a circle, $R_t\times S^1_{x\simeq x+2\pi R}$, with a constant $B$-field,
times an $R_\phi$ with a linear dilaton.

Now, following the same calculation done in (12)--(15) of~\cite{Chakraborty:2024ugc},
one finds that the energies $E_w(R)$ of winding $w$ long strings states with momentum $n$ on $S^1_x$,
in superstring theory on deformed BTZ, (\ref{defbtz})--(\ref{phiphi}),
with radial momentum $\sim j$ and in a state with transverse levels $N_{L,R}$, have the pattern:
\be\label{pattern}
E_w(R)+{wR\over\alpha'}(-1+b)=
E_w(R)+{wR\over\alpha'}\left(-1+\sqrt{\left(1+{\rho_-^2\over R^2}\right)\left(1+{\rho_+^2\over R^2}\right)}+b_0\right)
\ee
$$
=-{wR\over\alpha'}+\sqrt{{w^2R^2\over\alpha'^2}+{2\over\alpha'}\left(-{2j(j+1)\over k}+N_L+N_R-1\right)+{n^2\over R^2}}~.
$$
For completeness, we paste the calculation of~\cite{Chakraborty:2024ugc} in the appendix.

Next, following the same steps as in (16)--(22) of~\cite{Chakraborty:2024ugc}, one can argue that
\be\label{onefinds}
-{2j(j+1)\over k}+N_L+N_R-1={\cal E}_{\rm string}+{\rho_+^2+\rho_-^2\over 2\alpha'}~,
\ee
where ${\cal E}_{\rm string}$ is the (dimensionless) energy carried by a single string in the superstring on a BTZ black hole CFT, a.k.a.
\be\label{limit}
{\cal E}_{\rm string}=\lim_{{R\over\sqrt{\alpha'}}\to\infty}RE_1~,
\ee
and ${\rho_+^2+\rho_-^2\over 2\alpha'}$ is the contribution to the energy of a BTZ black hole (in units of $1/r_5$) due to a single $F1$
out of the $p$ that form it,
\be\label{mbtz}
{\rho_+^2+\rho_-^2\over 2\alpha'}={r_5M_{BTZ}\over p}~.
\ee
Equation (\ref{limit}) implies that, in the BTZ limit,
\be\label{btoone}
\lim_{{R\over\sqrt{\alpha'}}\to\infty}b=1~,
\ee
or, equivalently,
\be\label{bzerozero}
\lim_{{R\over\sqrt{\alpha'}}\to\infty}b_0\to 0~.
\ee
This amounts to the property of $B_{\tau\theta}^0$ in eq.~(\ref{restrict}).

From now on, we thus restrict to a $B$-field in eq.~(\ref{btautheta}) which satisfies eq.~(\ref{restrict}).
In this case, the pattern in eq.~(\ref{pattern}) can be written as
\be\label{patternn}
E_w(R)+{wR\over\alpha'}(-1+b)=
-{wR\over\alpha'}+\sqrt{{w^2R^2\over\alpha'^2}+{2\over\alpha'}\left({\cal E}_{\rm string}+{\rho_+^2+\rho_-^2\over 2\alpha'}\right)+{n^2\over R^2}}~.
\ee
Again, for completeness,~\footnote{And for presenting more properties, which do not appear explicitly above.}
we paste the calculation of~\cite{Chakraborty:2024ugc}~\footnote{Slightly modified (to include the {\it rotating} black-strings cases).}
in the appendix.

Now, we turn to single-trace $T\bar T$ holography.
First, define
\be\label{poslam}
\lambda\equiv{\alpha'\over R^2}~.
\ee
In~\cite{Chakraborty:2023zdd}, the following was shown.
The entropy, angular momentum and energy of the deformed BTZ black string above are
\be\label{s}
S={2\pi kp\over r_5}\tilde\rho_+~,
\ee
\be\label{n}
n_{F1}=J_{BTZ}={p\over\alpha'}\tilde\rho_+\tilde\rho_-~,
\ee
and~\footnote{See eqs.~(2.33), (2.32) with (2.22) and (2.34) in~\cite{Chakraborty:2023zdd}, respectively.}
\be\label{defe}
E_{\rm deformed\, BTZ}={Rp\over\alpha'}\left(-1+\sqrt{\left(1+{\tilde\rho_-^2\over R^2}\right)\left(1+{\tilde\rho_+^2\over R^2}\right)}\right)~,
\ee
respectively, where
\be\label{tilrho}
\tilde\rho_\pm\equiv{\rho_\pm\over\sqrt{1+{\rho_\mp^2\over R^2}}}~.
\ee
Then, on the $\lambda$-line of theories with fixed $\tilde\rho_\pm$,
\be\label{tilrhofix}
\tilde\rho_\pm(\lambda)\equiv{\rho_\pm(\lambda)\over\sqrt{1+{\rho_\mp^2(\lambda)\over R^2}}}=\tilde\rho_\pm(0)=\rho_\pm(0)~,
\ee
the black-strings microstates have the following properties:

\begin{itemize}

\item
Their number and momentum are $\lambda$-independent,
\be\label{snfix}
e^{S(\lambda)}=e^{S(0)}~,\qquad n_{F1}(\lambda)=n_{F1}(0)=J_{BTZ}~,
\ee
as follows from (\ref{s}) and (\ref{n}), respectively, with (\ref{tilrhofix}).

\item
Their deformed energy, (\ref{defe}), which we denote by
\be\label{elam}
E(\lambda)\equiv E_{\rm deformed\, BTZ}~,
\ee
satisfies
\be\label{ep}
{E(\lambda)\over p}={1\over\lambda R}\left(-1+\sqrt{1+2\lambda R{E(0)\over p}+\left({\lambda RP\over p}\right)^2}\right)~,
\ee
where
\be\label{ee}
E(0)\equiv{r_5\over R}M_{BTZ}~,\qquad P\equiv{J_{BTZ}\over R}~.
\ee

\item
The $p$'th fraction of the entropy (\ref{s}) can be written as
\be\label{ss}
{S\over p}=2\pi\sqrt{c\over 6}\left(\sqrt{E_L(1+\lambda E_R)}+\sqrt{E_R(1+\lambda E_L)}\right)~,\qquad c\equiv 6k~,
\ee
where
\be\label{eler}
E_{L,R}\equiv{R\over 2p}\left(E(\lambda)\pm P\right)~.
\ee

\end{itemize}
These properties are in harmony with single-trace $T\bar T$ holography, due to the following fact.~\footnote{See the discussion around (2.35)--(2.38) in~\cite{Chakraborty:2023zdd}, and more discussions concerning the recent status of single-trace $T\bar T$ holography in~\cite{Chakraborty:2023mzc}.}
The energy (\ref{ep}) and entropy (\ref{ss}) are the same as those of the states in a symmetric product ${\cal M}^p/S_p$,
where the seed theory ${\cal M}$ is a $CFT_2$ deformed by $T\bar T$ with a dimensionless deformation parameter $\lambda$, (\ref{poslam}),
whose energy and momentum are equally split among the $p$ copies of ${\cal M}$.

A couple of comments are in order:

\begin{itemize}

\item
The $1/p$ fraction of the energy of $\lambda$-deformed BTZ with mass $M_{BTZ}$ and angular momentum $J_{BTZ}$, (\ref{ep}),
can be written as
\be\label{ejbtz}
{E_{\rm deformed\,\,BTZ}(R)\over p}={1\over\lambda R}\left\{-1+\left[1-\left(\lambda J_{BTZ}\over p\right)^2\right]
\sqrt{\left(1+{\rho_-^2\over R^2}\right)\left(1+{\rho_+^2\over R^2}\right)}\right\}~.
\ee
One might thus be bothered by the large-$\lambda$ behavior of (\ref{ejbtz}), due to the sign change of
the $1-\left(\lambda J_{BTZ}\over p\right)^2$ factor, when $\lambda\geq p/J_{BTZ}$; this is addressed in the next item.

\item
Define
\be\label{qlqr}
q_L\equiv {J_{BTZ}\over R}-{pR\over\alpha'}~,\qquad q_R\equiv {J_{BTZ}\over R}+{pR\over\alpha'}~.
\ee
The relation between the fixed $\tilde\rho_\pm$ and $\rho_\pm\equiv\rho_\pm(\lambda)$ can now be written as
\be\label{canow}
\left({pR\over\alpha'}\right)^2\left(1+{\tilde\rho_\pm^2\over R^2}\right)=-q_Lq_R\left(1+{\rho_\pm^2\over R^2}\right)~.
\ee
Hence, $\rho_\pm\to\infty$ when $q_L\to 0$,~\footnote{Or when $q_R\to 0$, if $J_{BTZ}<0$.}
and we cannot continue to follow the state as we attempt to increase $\lambda\equiv{\alpha'/R^2}$ beyond $p/J_{BTZ}$
(recall that we consider $J_{BTZ}\geq 0$, w.l.g.).
Instead, we should consider the T-dual geometries of the above when the states become momentum dominated,
a.k.a. when ${J_{BTZ}\over R}>{pR\over\alpha'}$ (which is equivalent to $\lambda\geq p/J_{BTZ}$).
This addresses the concern in the previous item.

\end{itemize}

We are now ready to return to the long strings excitation in the deformed BTZ background.
The bottom line that follows from the above is the following.
If we choose~\footnote{Note that this $B_{\tau\theta}^0$ vanishes in the BTZ limit and/or when $\rho_-=0$,
and hence it doesn't contradict the arguments in~\cite{Martinec:2023plo} for BTZ (which amounts to $\lambda=0$),
and/or~\cite{Chakraborty:2023mzc} for global $AdS_3$ (which amounts to $\rho_-=0$ and $\rho_+^2=-r_5^2$; see e.g.~\cite{Chakraborty:2023mzc}).}
\be\label{choose}
B_{\tau\theta}^0\equiv B_{\tau\theta}(\rho=0)=
-{R^2\over r_5}{{\rho_-^2\rho_+^2\over R^4}\over\sqrt{\left(1+{\rho_-^2\over R^2}\right)\left(1+{\rho_+^2\over R^2}\right)}}~,
\ee
in~(\ref{btautheta}), namely, (\ref{bzero}),
\be\label{namely}
b_0=-{{\rho_-^2\rho_+^2\over R^4}\over\sqrt{\left(1+{\rho_-^2\over R^2}\right)\left(1+{\rho_+^2\over R^2}\right)}}~,
\ee
then, using
\be\label{using}
\sqrt{\left(1+{\rho_-^2\over R^2}\right)\left(1+{\rho_+^2\over R^2}\right)}
-{{\rho_-^2\rho_+^2\over R^4}\over\sqrt{\left(1+{\rho_-^2\over R^2}\right)\left(1+{\rho_+^2\over R^2}\right)}}
=\sqrt{\left(1+{\tilde\rho_-^2\over R^2}\right)\left(1+{\tilde\rho_+^2\over R^2}\right)}
\ee
(which follows from (\ref{tilrho})), we'll have, (\ref{binfty}), (\ref{b}),
\be\label{and}
b=\sqrt{\left(1+{\tilde\rho_-^2\over R^2}\right)\left(1+{\tilde\rho_+^2\over R^2}\right)}~,
\ee
and, (\ref{pattern}),
\be\label{well}
E_w(R)+{wR\over\alpha'}\left(-1+\sqrt{\left(1+{\tilde\rho_-^2\over R^2}\right)\left(1+{\tilde\rho_+^2\over R^2}\right)}\right)=
\ee
$$
=-{wR\over\alpha'}+\sqrt{{w^2R^2\over\alpha'^2}+{2\over\alpha'}\left(-{2j(j+1)\over k}+N_L+N_R-1\right)+{n^2\over R^2}}
$$
(which is the analog of eq. (15) in~\cite{Chakraborty:2024ugc} when $\rho_-\neq 0$ and with the $B$-field above),
and thus~\footnote{The reason for the tildes on the $\rho_\pm$ in ${\cal E}_{string}+{\tilde\rho_+^2+\tilde\rho_-^2\over 2\alpha'}$
is since we keep $-{2j(j+1)\over k}+N_L+N_R-1$ fixed,
and since it is equal to ${\cal E}_{string}+{\tilde\rho_+^2+\tilde\rho_-^2\over 2\alpha'}$
(the analog of (16) in~\cite{Chakraborty:2024ugc} when $\rho_-\neq 0$) at the BTZ point (since $\rho_\pm(\lambda=0)=\tilde\rho_\pm$),
it should be ${\cal E}_{string}+{\tilde\rho_+^2+\tilde\rho_-^2\over 2\alpha'}$ also when $\lambda\neq 0$
(to be fixed for all $\lambda$).}
\be\label{thus}
E_w(R)+{wR\over\alpha'}\left(-1+\sqrt{\left(1+{\tilde\rho_-^2\over R^2}\right)\left(1+{\tilde\rho_+^2\over R^2}\right)}\right)=
\ee
$$
=-{wR\over\alpha'}+\sqrt{{w^2R^2\over\alpha'^2}+{2\over\alpha'}\left({\cal E}_{string}+{\tilde\rho_+^2+\tilde\rho_-^2\over 2\alpha'}\right)+{n^2\over R^2}}
$$
(which is the analog of eq. (20) in~\cite{Chakraborty:2024ugc} when $\rho_-\neq 0$ and with the $B$-field above).

To recapitulate, the picture that emerges from (\ref{thus}) is the following.~\footnote{It is the generalization of that
summarized in section 3 of~\cite{Chakraborty:2024ugc}, for the long strings excitations in {\it rotating} black-strings backgrounds.}
First, let us denote the deformed BTZ black-hole (BH) energy,~(\ref{defe}), by~\footnote{To compactify the length of labels.}
\be\label{ebh}
E_{BH}\equiv E_{\rm deformed\, BTZ}~,
\ee
and the excitation energy of a long string, carrying winding $w$, by~\footnote{The `string' label is added
to emphasize the distinction between the excitation energy of the string as opposed to its contribution to the black-string energy.}
\be\label{ewstring}
E_{w\rm ,string}\equiv E_w~,
\ee
where $E_w$ is the one in (\ref{thus}).
Second, recall that the black-string background is formed by $p$ fundamental strings.
Now, consider the case of a single $w$-wound string excitation,
which amounts to the emission of a state formed by $w$ out of the $p$ strings that form the black string,
while the other $p-w$ strings are unexcited.
Each of these unexcited fundamental strings contributes equally to the energy of the deforemed BTZ black hole,
a $1/p$ fraction of its total energy,~(\ref{defe}),
\be\label{ebhp}
{E_{BH}R\over p}={1\over\lambda}\left(-1+\sqrt{\left(1+{\tilde\rho_-^2\over R^2}\right)\left(1+{\tilde\rho_+^2\over R^2}\right)}\right)~,
\qquad \lambda\equiv{\alpha'\over R^2}~.
\ee
On the other hand, the excited winding $w$ long string carries a total (dimensionless) energy,
\be\label{et}
E_{w\rm ,total}R=E_{w\rm ,string}R+w{E_{BH}R\over p}~.
\ee
The second term on the r.h.s. is the contribution of $w$ unexcited strings to the black-string energy,
while the first term on the r.h.s. is the additional energy due to exciting and emitting them from the black string,
in the form of a {\it single} winding $w$ long string.
Equation~(\ref{thus}) can now be written as
\be\label{thuswritten}
E_{w\rm ,total}R={w\over\lambda}\left(-1+\sqrt{1+{2\lambda{\cal E}_{w\rm ,total}\over w}+\left({\lambda n\over w}\right)^2}\right)~,
\qquad \lambda\equiv{\alpha'\over R^2}~,
\ee
where
\be\label{undefe}
{\cal E}_{w\rm ,total}={\cal E}_{w\rm ,string}+w{\sqrt{\alpha' k}M_{BTZ}\over p}~,
\ee
with
\be\label{eweone}
w\left({\cal E}_{w\rm ,string}+w{\sqrt{\alpha' k}M_{BTZ}\over p}\right)={\cal E}_{\rm string}+{\sqrt{\alpha' k}M_{BTZ}\over p}~,
\ee
for all $w=1,2,3,\dots$.~\footnote{Equation~(\ref{ebhp}) is a particular case of~(\ref{thuswritten})--(\ref{eweone}),
with $w=1$, ${\cal E}_{\rm string}=0$ and $n=J_{BTZ}/p$ (to see this, recall also the footnote near eq.~(\ref{thus})).}
Equation~(\ref{undefe}) is the undeformed total energy, obtained in the BTZ limit,
\be\label{btzlimit}
\lim_{{R\over\sqrt{\alpha'}}\to\infty}E_{w\rm ,total}R\to{\cal E}_{w\rm ,total}~.
\ee
Equation~(\ref{eweone}) implies that it is in harmony with the properties of a $w$-wound long string emitted from a BTZ black hole,~\cite{Martinec:2023plo} (see the appendix).
And, finally, eq.~(\ref{thuswritten}) implies that the deformed (dimensionless) energy spectrum, $E_{w\rm ,total}R$,
with undeformed spectrum ${\cal E}_{w\rm ,total}$,
satisfies precisely the $T\bar T$ formula of~\cite{Smirnov:2016lqw,Cavaglia:2016oda}, for a theory on a circle with radius $R$,
applied to the untwisted sector, if $w=1$, or to the $Z_w$ twisted sector, if $w>1$,
of a $p$-fold symmetric product of a $T\bar T$ deformed $CFT_2$ with $c=6k$.~\footnote{Recall
that the $w$-wound long-string sector corresponds to the $Z_w$ twisted sector of the symmetric orbifold; see e.g. the discussion around eqs.~(3.11)--(3.13) of~\cite{Giveon:2017myj}.}~\footnote{The block theory is associated with that of a single long string;
see e.g.~\cite{Chakraborty:2019mdf}, where it was denoted by ${\cal M}_{6k}^{(L)}$.
The central charge of the long-string $CFT_2$,~\cite{Seiberg:1999xz}, can be determined in various ways;
e.g. from~(\ref{elam})--(\ref{eler}) above.}

Hence, for the particular choice of the $B$-field,~(\ref{choose}), the long-strings spectrum is consistent with the picture that
a long-string excitation with $w=1$ deforms together with the black-string background as a block ${\cal M}$ of a symmetric product,
${\cal M}^p/S_p$,
while the energy of an emitted long string with $w>1$ deforms together with the background as a $Z_w$ twisted sector of a symmetric product,
in harmony with the single-trace $T\bar T$ holography conjecture
in~\cite{Giveon:2017nie,Chakraborty:2020swe,Chakraborty:2023mzc,Chakraborty:2023zdd}.

Comment:

It would be interesting to see if repeating what Emil Martinec argued in~\cite{Martinec:2023plo},~\footnote{See
below (2.11) in~\cite{Martinec:2023plo} and other places there, e.g. around (3.27)--(3.29).}
for deformed BTZ worldsheet CFT,~\footnote{E.g. one needs to extend (3.27)--(3.29) in~\cite{Martinec:2023plo}
to the change in entropy of the black string corresponding to our deformed BTZ theory upon emitting a winding string.}
and/or if arguments similar to those in section 4.2 of~\cite{Chakraborty:2024mls},
give rise to the $B_{\tau\theta}(\rho=0)$ in~(\ref{choose}), when $\lambda\neq 0$.~\footnote{It may be useful to recall that
${{\rho_-^2\rho_+^2\over R^4}\over\sqrt{\left(1+{\rho_-^2\over R^2}\right)\left(1+{\rho_+^2\over R^2}\right)}}=
\lambda^2{J_{BTZ}\over p}{\rho_-(\lambda)\rho_+(\lambda)\over\alpha'}=\left({\lambda J_{BTZ}\over p}\right)^2\left(1+{\cal O}(\lambda)\right)$,
and to recall (see comments above) that the geometry makes sense in string theory only when it's winding dominated,
a.k.a. when ${pR\over\alpha'}\geq{J_{BTZ}\over R}$
(otherwise, we should consider the T-dual configuration),
which implies that ${\lambda J_{BTZ}\over p}\leq 1$ (recall that we consider $J_{BTZ}\geq 0$, w.l.g.).}

\section{Negaive coupling}

In this section, we consider long-strings excitations in superstring theory on negatively deformed BTZ black holes,
formed by $p$ negative strings,~\cite{Chakraborty:2020swe,Chakraborty:2023zdd}.

The negatively deformed BTZ background appears in equations (3.10) with (3.11) of~\cite{Chakraborty:2023zdd}:
\be\label{ndefbtz}
ds^2=-{N^2\over 1-{\rho^2\over R^2}}d\tau^2+{d\rho^2\over N^2}+{\rho^2\over 1-{\rho^2\over R^2}}(d\theta-N_\theta d\tau)^2~,\qquad \theta\simeq\theta+2\pi~,
\ee
\be\label{nbtautheta}
B_{\tau\theta}=B_{\tau\theta}^0
+{\rho^2\over r_5}\sqrt{\left(1-{\rho_-^2\over R^2}\right)\left(1-{\rho_+^2\over R^2}\right)}{1\over 1-{\rho^2\over R^2}}~,
\qquad B_{\tau\theta}^0\equiv B_{\tau\theta}(\rho=0)~,
\ee
\be\label{ndilaton}
e^{2\Phi}={kv\over p}\sqrt{\left(1-{\rho_-^2\over R^2}\right)\left(1-{\rho_+^2\over R^2}\right)}{1\over 1-{\rho^2\over R^2}}~,
\qquad v\equiv {\rm Volume}(T^4)/\left(2\pi\sqrt{\alpha'}\right)^4~,
\ee
with
\be\label{nwithn}
N^2={\left(\rho^2-\rho_+^2\right)\left(\rho^2-\rho_-^2\right)\over r_5^2\rho^2}~, \qquad N_\theta={\rho_+\rho_-\over r_5\rho^2}~,
\qquad r_5\equiv\sqrt{\alpha' k}~.
\ee
This background describes a rotating black string in $2+1$ dimensions, whose properties are described in~\cite{Chakraborty:2023zdd}.

The comments and footnotes, which appear around eqs.~(\ref{nfone})--(\ref{restrict}), apply here as well.
In particular,
the value of $B_{\tau\theta}^0$ plays a central role also below,
and as argued in the previous section (and below), it should satisfy
\be\label{nrestrict}
B_{\tau\theta}^0(R/\sqrt{\alpha'}\to\infty)\to 0~,
\ee
for consistency with the BTZ limit.

Next, define,
\be\label{ntx}
t\equiv{R\tau\over r_5}~,\qquad x\equiv R\theta~,\qquad{\phi\over r_5}\equiv\log\left(\rho\over R\right)~,
\ee
and
\be\label{nbzero}
b_0\equiv B_{tx}(\rho=0)={r_5\over R^2}B_{\tau\theta}^0~,
\ee
as well as
\be\label{nbinfty}
b\equiv B_{tx}(\rho\to\infty)={r_5\over R^2}B_{\tau\theta}(\rho\to\infty)~.
\ee
Asymptotically, a.k.a. at $\rho/R\to\infty$, the black-string background~(\ref{ndefbtz})--(\ref{nwithn})  takes the form
\be\label{nds}
ds^2\to dt^2-dx^2+d\phi^2~,
\ee
\be\label{nb}
B_{tx}\to b=b_0-\sqrt{\left(1-{\rho_-^2\over R^2}\right)\left(1-{\rho_-^2\over R^2}\right)}~,
\ee
\be\label{nphiphi}
\Phi\to -{\phi\over r_5}+const~.
\ee
Due to the signature flip in the $(t,x)$ directions in~(\ref{nds}) (relative to~(\ref{ds})), the asymptotic background is a flat spacetime
consisting of a {\it compact time}, $S^1_{x\simeq x+2\pi R}$, times a non-compact space, $R_t$, with a constant $B$-field,~(\ref{nb}),
times an $R_\phi$ with a linear dilaton.

It is important to emphasize the following:
\begin{itemize}
\item
Even though, asymptotically, now $x$ is timelike while $t$ is spacelike,
in the holographic dictionary, we regard $t$ as time, and its conjugate momentum, $E$, as energy; see~\cite{Chakraborty:2020swe} for detailed
considerations of these properties.
\item
Note also the sign differences in the $b$ of~(\ref{nb}), relative to the positive case in~(\ref{b}); these will play a role below.
\item
The `$const$' in the asymptotic behavior of the linear dilaton,~(\ref{nphiphi}), is complex; this will not play a role below.
\item
Although, naively, the above seem to be pathological,
we expect string theory on these negatively deformed BTZ worldsheet CFT's to be consistent theories;
see discussions on that in~\cite{Chakraborty:2020swe,Chakraborty:2023zdd}.
\end{itemize}

Now, following similar calculations as for the positive case (see the appendix), one finds
\be\label{npattern}
E_w(R)+{wR\over\alpha'}(1+b)=
E_w(R)+{wR\over\alpha'}\left(1-\sqrt{\left(1-{\rho_-^2\over R^2}\right)\left(1-{\rho_+^2\over R^2}\right)}+b_0\right)=
\ee
$$
={wR\over\alpha'}-\sqrt{{w^2R^2\over\alpha'^2}-{2\over\alpha'}\left(-{2j(j+1)\over k}+N_L+N_R-1\right)+{n^2\over R^2}}
$$
(instead of~(\ref{pattern})).

Next, following the same arguments around eqs.~(\ref{onefinds})--(\ref{mbtz})
(and their details around eqs.~(\ref{h})--(\ref{de1}) in the appendix),
one finds that in the BTZ limit,
\be\label{nbtoone}
\lim_{{R\over\sqrt{\alpha'}}\to\infty}b=-1
\ee
(as opposed to (\ref{btoone})), or, equivalently,
\be\label{nbzerozero}
\lim_{{R\over\sqrt{\alpha'}}\to\infty}b_0\to 0
\ee
(as in~(\ref{bzerozero})),
and we thus restrict to a $B$-field in eq.~(\ref{nbtautheta}) which satisfies eq.~(\ref{nrestrict}), as before.
In this case, using~(\ref{onefinds}),
the pattern in eq.~(\ref{npattern}) can be written as
\be\label{npatternn}
E_w(R)+{wR\over\alpha'}(1+b)=
{wR\over\alpha'}-\sqrt{{w^2R^2\over\alpha'^2}-{2\over\alpha'}\left({\cal E}_{\rm string}+{\rho_+^2+\rho_-^2\over 2\alpha'}\right)+{n^2\over R^2}}
\ee
(instead of~(\ref{patternn})).

Now, we turn to single-trace $T\bar T$ holography.
First, define
\be\label{nposlam}
\lambda\equiv -{\alpha'\over R^2}<0
\ee
(as opposed to~(\ref{poslam})).
In~\cite{Chakraborty:2023zdd}, the following was shown.
The entropy, angular momentum and energy of the deformed BTZ black string above are given by~(\ref{s}),~(\ref{n})
and~\footnote{See eqs.~(2.33), (2.32) with (2.22) and (3.22) in~\cite{Chakraborty:2023zdd}, respectively.}
\be\label{ndefe}
E_{\rm deformed\, BTZ}={Rp\over\alpha'}\left(1-\sqrt{\left(1-{\tilde\rho_-^2\over R^2}\right)\left(1-{\tilde\rho_+^2\over R^2}\right)}\right)~,
\ee
respectively, where
\be\label{ntilrho}
\tilde\rho_\pm\equiv{\rho_\pm\over\sqrt{1-{\rho_\mp^2\over R^2}}}~.
\ee
Then, on the $\lambda$-line of theories with fixed $\tilde\rho_\pm$,
\be\label{ntilrhofix}
\tilde\rho_\pm(\lambda)\equiv{\rho_\pm(\lambda)\over\sqrt{1-{\rho_\mp^2(\lambda)\over R^2}}}=\tilde\rho_\pm(0)=\rho_\pm(0)~,
\ee
the black-strings microstates have the following properties:

\begin{itemize}

\item
Their number and momentum are $\lambda$-independent,~(\ref{snfix}),
as follows from (\ref{s}) and (\ref{n}), respectively, with (\ref{ntilrhofix}).

\item
Their deformed energy, (\ref{ndefe}), is given by~(\ref{ep}), (\ref{ee}) with~(\ref{elam}),
where now $\lambda<0$ (the one given in~(\ref{nposlam}), instead of~(\ref{poslam}) in the previous section).~\footnote{It
can be written also in the form~(\ref{ejbtz}), with the $\lambda<0$ in~(\ref{nposlam}).}

\item
The $p$'th fraction of the entropy (\ref{s}) can be written as~(\ref{ss}) with~(\ref{eler}),
with the negative $\lambda$ in~(\ref{nposlam}) (instead of~(\ref{poslam}) in the previous section).

\end{itemize}
These properties are in harmony with {\it negative single-trace $T\bar T$ holography},
for the same reasons as those below~(\ref{eler}), but now for a $CFT_2$ deformed by $T\bar T$
with a negative dimensionless deformation parameter $\lambda<0$, (\ref{nposlam}).~\footnote{See
the discussion in the end of section 3 of~\cite{Chakraborty:2023zdd}.}

We are now ready to return to the long strings excitation in the deformed BTZ background.
The bottom line that follows from the above is the following.
If we choose~\footnote{As in the $\lambda>0$ case,
note that this $B_{\tau\theta}^0$ vanishes in the BTZ limit and/or when $\rho_-=0$,
and hence it doesn't contradict the arguments in~\cite{Martinec:2023plo} for BTZ (which amounts to $\lambda=0$),
and/or~\cite{Chakraborty:2023mzc} for global $AdS_3$ (which amounts to $\rho_-=0$ and $\rho_+^2=-r_5^2$; see e.g.~\cite{Chakraborty:2023mzc}),
regardless of the sign of $\lambda$. Moreover, $B_{\tau\theta}^0\to 0$ when $\rho_+\to R$ for any finite momentum $n_{F1}$
(which is given by~(\ref{n}) with~(\ref{ntilrho}) when $\lambda<0$) that the negative strings carry
(since $\rho_-\to 0$ in the $\rho_+\to R$ limit); hence, the discussion in~\cite{Giveon:2023rsk} is not affected by the different choice
of $B_{\tau\theta}^0$ here (it was chosen to be zero there also when $\rho_-\neq 0$).}
\be\label{nchoose}
B_{\tau\theta}^0\equiv B_{\tau\theta}(\rho=0)=
{R^2\over r_5}{{\rho_-^2\rho_+^2\over R^4}\over\sqrt{\left(1-{\rho_-^2\over R^2}\right)\left(1-{\rho_+^2\over R^2}\right)}}~,
\ee
in~(\ref{nbtautheta}), namely, (\ref{nbzero}),
\be\label{nnamely}
b_0={{\rho_-^2\rho_+^2\over R^4}\over\sqrt{\left(1-{\rho_-^2\over R^2}\right)\left(1-{\rho_+^2\over R^2}\right)}}~,
\ee
then, using
\be\label{nusing}
\sqrt{\left(1-{\rho_-^2\over R^2}\right)\left(1-{\rho_+^2\over R^2}\right)}
-{{\rho_-^2\rho_+^2\over R^4}\over\sqrt{\left(1-{\rho_-^2\over R^2}\right)\left(1-{\rho_+^2\over R^2}\right)}}
=\sqrt{\left(1-{\tilde\rho_-^2\over R^2}\right)\left(1-{\tilde\rho_+^2\over R^2}\right)}
\ee
(which follows from (\ref{ntilrho})), we'll have, (\ref{nbinfty}), (\ref{nb}),
\be\label{nand}
b=-\sqrt{\left(1-{\tilde\rho_-^2\over R^2}\right)\left(1-{\tilde\rho_+^2\over R^2}\right)}~,
\ee
and, (\ref{npattern}),
\be\label{nwell}
E_w(R)+{wR\over\alpha'}\left(1-\sqrt{\left(1-{\tilde\rho_-^2\over R^2}\right)\left(1-{\tilde\rho_+^2\over R^2}\right)}\right)=
\ee
$$
={wR\over\alpha'}-\sqrt{{w^2R^2\over\alpha'^2}-{2\over\alpha'}\left(-{2j(j+1)\over k}+N_L+N_R-1\right)+{n^2\over R^2}}
$$
(which is the analog of eq. (15) in~\cite{Chakraborty:2024ugc} when $\rho_-\neq 0$, $\lambda<0$, and with the $B$-field above),
and thus~\footnote{As in the $\lambda>0$ case,
the reason for the tildes on the $\rho_\pm$ in ${\cal E}_{string}+{\tilde\rho_+^2+\tilde\rho_-^2\over 2\alpha'}$
is since we keep $-{2j(j+1)\over k}+N_L+N_R-1$ fixed,
and since it is equal to ${\cal E}_{string}+{\tilde\rho_+^2+\tilde\rho_-^2\over 2\alpha'}$
(the analog of (16) in~\cite{Chakraborty:2024ugc} when $\rho_-\neq 0$) at the BTZ point (since $\rho_\pm(\lambda=0)=\tilde\rho_\pm$),
it should be ${\cal E}_{string}+{\tilde\rho_+^2+\tilde\rho_-^2\over 2\alpha'}$ also when $\lambda\neq 0$
(to be fixed for all $\lambda$, regardless of its sign).}
\be\label{nthus}
E_w(R)+{wR\over\alpha'}\left(1-\sqrt{\left(1-{\tilde\rho_-^2\over R^2}\right)\left(1-{\tilde\rho_+^2\over R^2}\right)}\right)=
\ee
$$
={wR\over\alpha'}-\sqrt{{w^2R^2\over\alpha'^2}-{2\over\alpha'}\left({\cal E}_{string}+{\tilde\rho_+^2+\tilde\rho_-^2\over 2\alpha'}\right)+{n^2\over R^2}}
$$
(which is the analog of eq. (20) in~\cite{Chakraborty:2024ugc} when $\rho_-\neq 0$, $\lambda<0$, and with the $B$-field above).

To recapitulate, the picture that emerges from (\ref{nthus}) is similar to that in the $\lambda>0$ case.~\footnote{It
is the generalization of that summarized in section 3 of~\cite{Chakraborty:2024ugc}, for the long strings excitations in {\it rotating} black-strings backgrounds
obtained by deforming BTZ black holes in the {\it negative} direction, $\lambda<0$.}
To describe it, we'll use the same notation as in~(\ref{ebh}) and (\ref{ewstring}),
and recall that the black-string background in eqs.~(\ref{ndefbtz})--(\ref{nwithn})
is formed by $p$ {\it negative strings},~\cite{Chakraborty:2020swe,Chakraborty:2023zdd}.~\footnote{See
e.g. below (3.7) in~\cite{Chakraborty:2023zdd}.}
Now, repeating word by word the discussion from the positive case,~\footnote{With tiny modifications, adapted to the negative case here.}
consider the case of a single $w$-wound string excitation,
which amounts to the emission of a state formed by $w$ out of the $p$ negative strings that form the black string,
while the other $p-w$ negative strings are unexcited.
Each of these unexcited negative strings contributes equally to the energy of the deforemed BTZ black hole,
a $1/p$ fraction of its total energy,~(\ref{ndefe}),
\be\label{nebhp}
{E_{BH}R\over p}={1\over\lambda}\left(-1+\sqrt{\left(1-{\tilde\rho_-^2\over R^2}\right)\left(1-{\tilde\rho_+^2\over R^2}\right)}\right)~,
\qquad \lambda\equiv -{\alpha'\over R^2}~.
\ee
On the other hand, the excited winding $w$ long string carries a total (dimensionless) energy,
\be\label{net}
E_{w\rm ,total}R=E_{w\rm ,string}R+w{E_{BH}R\over p}~.
\ee
The second term on the r.h.s. is the contribution of $w$ unexcited negative strings to the black-string energy,
while the first term on the r.h.s. is the additional energy due to exciting and emitting them from the black string,
in the form of a {\it single} winding $w$ long string.
Equation~(\ref{nthus}) can now be written as
\be\label{nthuswritten}
E_{w\rm ,total}R={w\over\lambda}\left(-1+\sqrt{1+{2\lambda{\cal E}_{w\rm ,total}\over w}+\left({\lambda n\over w}\right)^2}\right)~,
\qquad \lambda\equiv -{\alpha'\over R^2}~,
\ee
where
\be\label{nundefe}
{\cal E}_{w\rm ,total}={\cal E}_{w\rm ,string}+w{\sqrt{\alpha' k}M_{BTZ}\over p}~,
\ee
with
\be\label{neweone}
w\left({\cal E}_{w\rm ,string}+w{\sqrt{\alpha' k}M_{BTZ}\over p}\right)={\cal E}_{\rm string}+{\sqrt{\alpha' k}M_{BTZ}\over p}~,
\ee
for all $w=1,2,3,\dots$.~\footnote{Equation~(\ref{nebhp}) is a particular case of~(\ref{nthuswritten})--(\ref{neweone}),
with $w=1$, ${\cal E}_{\rm string}=0$ and $n=J_{BTZ}/p$ (to see this, recall also the footnote near eq.~(\ref{nthus})).}
Equation~(\ref{nundefe}) is the undeformed total energy, obtained in the BTZ limit,
\be\label{nbtzlimit}
\lim_{{R\over\sqrt{\alpha'}}\to\infty}E_{w\rm ,total}R\to{\cal E}_{w\rm ,total}~.
\ee
Equation~(\ref{neweone}) implies that it is in harmony with the properties of a $w$-wound long string emitted from a BTZ black hole,~\cite{Martinec:2023plo} (see the appendix).
And, finally, eq.~(\ref{nthuswritten}) implies that the deformed (dimensionless) energy spectrum, $E_{w\rm ,total}R$,
with undeformed spectrum ${\cal E}_{w\rm ,total}$,
satisfies precisely the $T\bar T$ formula of~\cite{Smirnov:2016lqw,Cavaglia:2016oda}, for a theory on a circle with radius $R$,
applied to the untwisted sector, if $w=1$, or to the $Z_w$ twisted sector, if $w>1$,
of a $p$-fold symmetric product of a $T\bar T$ deformed $CFT_2$ with $c=6k$.~\footnote{As in the positive case, recall
that the $w$-wound long-string sector corresponds to the $Z_w$ twisted sector of the symmetric orbifold; see e.g. the discussion around eqs.~(3.11)--(3.13) of~\cite{Giveon:2017myj}.}~\footnote{The block theory is associated with that of a single negative long string;
see e.g.~\cite{Chakraborty:2019mdf}, where it was denoted by ${\cal M}_{6k}^{(L)}$.
As in the positive case, the central charge of the long-string $CFT_2$,~\cite{Seiberg:1999xz}, can be determined in various ways;
e.g. from~(\ref{elam})--(\ref{eler}) above (which apply, as we found, to {\it any} $\lambda$, regardless of its sign).}

Hence, with the particular choice of the $B$-field,~(\ref{nchoose}),
for the $\lambda<0$ case (as in the $\lambda>0$ case with (\ref{choose})),
the long-strings spectrum is consistent with the picture that a long-string excitation with $w=1$
deforms together with the black-string background as a block ${\cal M}$ of a symmetric product, ${\cal M}^p/S_p$,
while the energy of an emitted long string with $w>1$ deforms together with the background as a $Z_w$ twisted sector of a symmetric product,
in harmony with the negative single-trace $T\bar T$ holography conjecture
in~\cite{Giveon:2017nie,Chakraborty:2020swe,Chakraborty:2023mzc,Chakraborty:2023zdd}.

\section{Other $B_{\tau\theta}(\rho=0)$}

In the previous sections, we found that in single-trace $T\bar T$ holography, the
\be\label{bbbzero}
b_0\equiv{r_5\over R^2}B_{\tau\theta}(\rho=0)
\ee
(and
\be\label{bbbb}
b\equiv {r_5\over R^2}B_{\tau\theta}(\rho\to\infty)=b_0\pm\sqrt{\left(1\pm{\rho_-^2\over R^2}\right)\left(1\pm{\rho_+^2\over R^2}\right)}~,
\ee
correspondingly) is:
\be\label{btt}
b_0^{T\bar T}=\mp{{\rho_-^2\rho_+^2\over R^4}\over\sqrt{\left(1\pm{\rho_-^2\over R^2}\right)\left(1\pm{\rho_+^2\over R^2}\right)}}\qquad
\left(b^{T\bar T}=\pm\sqrt{\left(1\pm{\tilde\rho_-^2\over R^2}\right)\left(1\pm{\tilde\rho_+^2\over R^2}\right)}\right)~,
\ee
where the $T\bar T$ label in~(\ref{btt}) is added for convenience below,
and the $\pm$'s or $\mp$ in~(\ref{bbbb}) and/or~(\ref{btt}) amount to the positive and negative single-trace $T\bar T$ holography, respectively.

In this section, we inspect a couple of other choices that appear in the literature.

Consider first the superstring on the deformed BTZ black-holes backgrounds studied in~\cite{Apolo:2019zai,Apolo:2021wcn}:~\footnote{See
(3.14) in ~\cite{Apolo:2021wcn} (with $\lambda_\pm=0$).}
\be\label{asds}
{ds^2\over r_5^2}={dr^2\over 4\left(r^2-4T_u^2T_v^2\right)}+{rdudv+T_u^2du^2+T_v^2dv^2\over 1+\lambda_0 r+\lambda_0^2T_u^2T_v^2}~,
\ee
\be\label{asb}
{B\over r_5^2}={r+2\lambda_0T_u^2T_v^2\over 1+\lambda_0 r+\lambda_0^2T_u^2T_v^2}{dv\wedge du\over 2}~,
\qquad e^{2\Phi}={kv\over p}{1-\lambda_0^2T_u^2T_v^2\over 1+\lambda_0 r+\lambda_0^2T_u^2T_v^2}~.
\ee
Defining
\be\label{tpm}
T_\pm\equiv T_u\pm T_v~,
\ee
and performing the coordinates transformation
\be\label{asruv}
r={\rho^2\over r_5^2}-{T_+^2+T_-^2\over 2}~,\qquad
u={R\sqrt{\lambda_0}\over r_5}\left(\theta-{\tau\over r_5}\right)~,\qquad v={R\sqrt{\lambda_0}\over r_5}\left(\theta+{\tau\over r_5}\right)~,
\ee
as well as the parameters identification
\be\label{asrt}
{R^2\over r_5^2}=\pm{\left(1-\lambda_0T_u^2\right)\left(1-\lambda_0T_v^2\right)\over\lambda_0}~,
\qquad T_\pm={\rho_\pm\over r_5}~,
\ee
one finds the backgrounds in~(\ref{defbtz})--(\ref{withn}) and in~(\ref{ndefbtz})--(\ref{nwithn}),
for the $\pm$ in~(\ref{asrt}), respectively,
with
\be\label{asbzero}
B_{\tau\theta}^{ADS}(\rho=0)=\pm{R^2\over r_5}\left(1-\sqrt{\left(1\pm{\rho_-^2\over R^2}\right)\left(1\pm{\rho_+^2\over R^2}\right)}\right)~,
\ee
where the ADS labels the authors of~\cite{Apolo:2019zai,Apolo:2021wcn},~\footnote{The relations
$$
1+\lambda_0 r+\lambda_0^2T_u^2T_v^2=\pm{\lambda_0R^2\over r_5^2}\left(1\pm{\rho^2\over R^2}\right)~,
$$
$$
{1\over\lambda_0}=\pm{R^2\over 2r_5^2}\left(1+\sqrt{\left(1\pm{\rho_-^2\over R^2}\right)\left(1\pm{\rho_+^2\over R^2}\right)}\right)+{\rho_+^2+\rho_-^2\over 4r_5^2}~,
$$
$$
\lambda_0={4r_5^2\left(\rho_+^2+\rho_-^2\right)\over\left(\rho_+^2-\rho_-^2\right)^2}
\left[1\pm{2R^2\over\rho_+^2+\rho_-^2}\left(1-\sqrt{\left(1\pm{\rho_-^2\over R^2}\right)\left(1\pm{\rho_+^2\over R^2}\right)}\right)\right]~,
$$
which follow from~(\ref{asrt}), are useful.}
and thus
\be\label{bads}
b_0^{ADS}=\pm\left(1-\sqrt{\left(1\pm{\rho_-^2\over R^2}\right)\left(1\pm{\rho_+^2\over R^2}\right)}\right)
\qquad \left(b^{ADS}=\pm 1\right)~.
\ee

Comment:

Note that
\be\label{adslim}
\lim_{R\to\infty}B^{ADS}_{\tau\theta}(\rho=0)=\lim_{R\to\infty}{R^2\over r_5}b^{ADS}_0=-{\rho_+^2+\rho_-^2\over 2r_5}~,
\ee
in tension with the BTZ limit,~(\ref{restrict}),~(\ref{nrestrict}) (unless $\rho_\pm=0$, the massless BTZ case).
Moreover, it's in dissonance with single-trace $T\bar T$ holography,~(\ref{btt}) (again, apart from the massless BTZ case).

In~\cite{Chakraborty:2023mzc,Chakraborty:2023zdd}, the choice was
\be\label{bcgk}
b_0^{CGK}=0\qquad \left(b^{CGK}=\pm\sqrt{\left(1\pm{\rho_-^2\over R^2}\right)\left(1\pm{\rho_+^2\over R^2}\right)}\right)~,
\ee
where the CGK labels the authors of~\cite{Chakraborty:2023mzc,Chakraborty:2023zdd}.

Comment:

Recall that although the CGK coice, $B^{CGK}_{\tau\theta}(\rho=0)=0$, is manifestly
in harmony with its correct value in the BTZ limit,~(\ref{restrict}),~(\ref{nrestrict}),
it is equal to its value in single-trace $T\bar T$  holography,~(\ref{btt}),
only in the {\it non}-rotating black-strings cases, $\rho_-=0$.

Our last example is deformed rotating BTZ backgrounds obtained by boosting the non-rotating ones.
Concretely, we start from the backgrounds in~(\ref{defbtz})--(\ref{withn}) (or~(\ref{ndefbtz})--(\ref{nwithn}))
with $\rho_-=0$, and with
\be\label{brrzero}
B_{\tau\theta}(\rho=\rho_-=0)=0
\ee
(which is the value compatible with single-trace $T\bar T$ holography,~(\ref{btt}), when $\rho_-=0$),
and rewrite them in terms of
\be\label{ttxx}
t\equiv{R\tau\over r_5}~,\qquad x\equiv R\theta~,
\ee
for which the line element in the $(t,x)$ direction is canonically normalized at $\rho\to\infty$,
\be\label{dstx}
ds^2_{(t,x)}\to -dt^2+dx^2~.
\ee
Then, we generate the rotating black-string backgrounds by boosting the non-rotating ones in the $x$ direction,
with rapidity $\alpha$ given by
\be\label{tanha}
\tanh\alpha={\rho_-\over\rho_+}~,
\ee
followed by the coordinate transformation
\be\label{cortrans}
\rho^2\to\cosh^2\alpha\left(\rho^2-\rho_-^2\right)~,
\ee
with the $R$ of~(\ref{defbtz})--(\ref{withn}) and~(\ref{ndefbtz})--(\ref{nwithn}) at the point $\alpha$,
denoted by $R_\alpha$ below, obeying
\be\label{obeying}
{R_\alpha^2\cosh^2\alpha\over R_0^2}={1\over 1\pm{\rho_-^2\over R_\alpha^2}}~,
\ee
where the $\pm$ amounts to~(\ref{defbtz})--(\ref{withn}) and~(\ref{ndefbtz})--(\ref{nwithn}), respectively.
We find the solutions in~(\ref{defbtz})--(\ref{withn}) and~(\ref{ndefbtz})--(\ref{nwithn}),~\footnote{Up to a shift of the dilaton.}
with
\be\label{shifting}
B_{\tau\theta}^{\rm boost}(\rho=0)=-{\rho_-^2\over r_5}\sqrt{1\pm{\rho_+^2\over R^2}\over 1\pm{\rho_-^2\over R^2}}~,
\ee
where here we erased the $\alpha$ label from the $R_\alpha$ in~(\ref{obeying}).

To recapitulate,
in the rotating deformed BTZ black holes obtained by boosting the non-rotating backgrounds ((\ref{defbtz})--(\ref{withn}) and~(\ref{ndefbtz})--(\ref{nwithn}) with $\rho_-=0$ and $B_{\tau\theta}(\rho=0)=0$),
we have
\be\label{bboost}
b_0^{\rm boost}=-{\rho_-^2\over R^2}\sqrt{1\pm{\rho_+^2\over R^2}\over 1\pm{\rho_-^2\over R^2}}\qquad
\left(b^{\rm boost}=\pm\sqrt{1\pm{\rho_+^2\over R^2}\over 1\pm{\rho_-^2\over R^2}}\right)~.
\ee
A couple of comments are in order:

\begin{itemize}

\item
Note that
\be\label{boostlim}
\lim_{R\to\infty}B^{\rm boost}_{\tau\theta}(\rho=0)=\lim_{R\to\infty}{R^2\over r_5}b^{\rm boost}_0=-{\rho_-^2\over r_5}~,
\ee
in tension with the BTZ limit~(\ref{restrict}),~(\ref{nrestrict}) (unless $\rho_-=0$, the non-rotating BTZ black holes
cases~\footnote{As well as global $AdS_3$, which amounts to $\rho_+^2=-r_5^2$ and $\rho_-=0$ (see e.g.~\cite{Chakraborty:2023mzc}).}).

\item
In appendix C of~\cite{Giveon:2005mi}, black-string solutions were obtained as the sigma-model backgrounds of
$(SL(2,R)\times U(1))/U(1)$ coset CFTs.
It can be shown~\footnote{We shall not present the coordinates transformation and parameters identification here.}
that these backgrounds are identical to those in (\ref{defbtz})--(\ref{withn}),
with the $B$-field of the boosted solution above,~(\ref{shifting}),~\footnote{The fact that in the coset backgrounds
$B_{\tau\theta}(\rho=0)\to -\rho_-^2/r_5$ in the BTZ limit, $\lambda\equiv\alpha'/R^2\to 0$, implies also that they do not produce the choice
$B_{\tau\theta}(\rho=0)=0$, when $\rho_-\neq 0$, which was argued to be the correct one for BTZ,~\cite{Martinec:2023plo} (see the appendix).
(One possibility is that we were not careful about invariance under large gauge transformations,
as we were in~\cite{Giveon:1991jj}, in the asymmetric case, a.k.a. when $\psi\neq 0$ in~\cite{Giveon:2005mi},
which amounts to $\rho_-\neq 0$.
Another possibility is that both cases amount to consistent worldsheet CFTs).}
\be\label{bcoset}
B_{\tau\theta}^{\rm coset}(\rho=0)=-{\rho_-^2\over r_5}\sqrt{1+{\rho_+^2\over R^2}\over 1+{\rho_-^2\over R^2}}~.
\ee

\item
The fact that turning on momentum is obtained by boosting the non-rotating solutions,
as is done above, propagates from the fact that the backgrounds corresponding
to $p$ fundamental strings, $F1$, with momentum $n_{F1}=J_{BTZ}$ on $S^1$, in the presence of fivebranes, $NS5$,
on that $S^1$, is obtained by boosting the zero-momentum solutions.
The black-strings backgrounds above amount to the near $NS5$ regime
and the BTZ limit is obtained in the near $F1$ theory.~\footnote{See details in~\cite{Chakraborty:2020swe,Chakraborty:2023zdd}
and references therein.}

\end{itemize}

\vspace{10mm}

\section*{Acknowledgments}
We thank Soumangsu Chakraborty for discussions.
This work was supported in part by the ISF (grant number 256/22).

\vspace{10mm}

\section*{Appendix -- Calculations from \cite{Chakraborty:2024ugc} and (\ref{pattern}),~(\ref{patternn}),~(\ref{npattern})}

A large class of observables in the (NS,NS) sector of superstring theory on deformed BTZ black holes is given by vertex operators
in the $(-1,-1)$ picture, whose behavior in the asymptotically linear dilaton regime,~(\ref{ds})--(\ref{phiphi}), is
\be\label{v}
V_{phys}\to e^{-\varphi-\bar\varphi}V_{N_L,N_R}e^{-iE t}e^{ip_Lx_L+ip_Rx_R}e^{2j\phi/r_5}~,
\ee
where
\be\label{plpr}
(p_L,p_R)=\left({wR\over\alpha'}+{n\over R},{wR\over\alpha'}-{n\over R}\right)~,
\ee
and
\be\label{onshell}
{\alpha'\over 4}\left(-(E+bwR/\alpha')^2+p_{L,R}^2\right)-{j(j+1)\over k}+N_{L,R}={1\over 2}~.
\ee
Each of these operators amounts to a string with winding $w$ and momentum $n$ on the asymptotic circle, $S^1_{x\simeq x+2\pi R}$,
moving with momentum governed by the quantum number $j$ in the radial direction,
in a particular state of transverse left and right-handed levels, $N_{L,R}$;
eq.~(\ref{onshell}) is obtained from the mass-shell relation of physical vertex operators
and using e.g. (2.4.11) in~\cite{Giveon:1994fu}~\footnote{More details:
for canonically normalized time, $t$,
times a circle, parameterized by $y\sim y+2\pi$,
with radius $r_y$ in $\sqrt{\alpha'}$ units,
manipulations in~\cite{Giveon:1994fu}, adapted to time times a circle, give
$$
\Delta+\bar\Delta={1\over 2}\left(n_i(g^{-1})^{ij}n_j+m^i(g-bg^{-1}b)_{ij}m^j+2m^ib_{ik}(g^{-1})^{kj}n_j\right)
$$
with
$$
g_{ij}=\pmatrix{
 -1&0\cr
 0&r_y^2&
},\qquad
b_{ij}=\pmatrix{
0 & -br_y\cr
br_y& 0
}
$$
$$
n_i=(e,n_y)~,\qquad m^i=(0,w_y)~,\qquad i,j=t,y
$$
for (dimensionless) energy $e\equiv\sqrt{\alpha'}E$ and momentum and winding $n_y,w_y$ on $S^1_y$,
from which one finds
$$
\Delta+\bar\Delta={1\over 2}\left(-(e+br_yw_y)^2+{n_y^2\over r_y^2}+w_y^2r_y^2\right)~,\qquad \Delta-\bar\Delta=n_yw_y
$$
in harmony with the first term in~(\ref{onshell}).
}
(adapted to~(\ref{ds})).

From eq.~(\ref{onshell}), we now get~\footnote{The label $w$ is added to the $E$ above for convenience later.}
\be\label{e}
E_w+{wR\over\alpha'}(-1+b)=
-{wR\over\alpha'}+\sqrt{{w^2R^2\over\alpha'^2}+{2\over\alpha'}\left(-{2j(j+1)\over k}+N_L+N_R-1\right)+{n^2\over R^2}}~.
\ee
This is the result in eq.~(\ref{pattern}).

For long strings with radial momentum ${2s\over r_5}$, a.k.a. with $j=-{1\over 2}+is$, $s\in R$,
we can keep $j$ and all the other quantum numbers fixed, as we vary $R$, thus obtaining a flow of energies, $E_w(R)$,
for each winding number, $w=1,2,3,\dots$, in eq.~(\ref{e}).

The undeformed BTZ background and the long strings spectrum of the superstring on BTZ are obtained by considering the background~(\ref{defbtz})--(\ref{withn})
in the limit $\rho_\pm,\rho\ll R$,~\footnote{See eqs. (2.20)--(2.23) and the discussion around them in~\cite{Chakraborty:2023zdd}.}
and taking the $R/\sqrt{\alpha'}\to\infty$ limit of~(\ref{e}), respectively.
Now, following~\cite{Martinec:2023plo} and references therein, for long strings in BTZ,~\footnote{See,
in particular, the discussion around eqs. (2.32)--(2.34) in~\cite{Martinec:2023plo}.}
\be\label{h}
-{2j(j+1)\over k}+N_L+N_R-1=w\left({\cal E}_{w,\rm string}+{w\left(\rho_+^2+\rho_-^2\right)\over 2\alpha'}\right)~,
\ee
where ${\cal E}_{w,\rm string}$ is the (dimensionless) excitation energy carried by a string carrying winding $w$
in the superstring on BTZ,~\footnote{See eq. (2.34) in~\cite{Martinec:2023plo}.}
and ${w\left(\rho_+^2+\rho_-^2\right)\over 2\alpha'}$ is the contribution to the energy of a BTZ black hole (in units of $1/r_5$)
due to $w$ fundamental strings on the $\theta$ direction of the BTZ limit of~(\ref{defbtz}),
out of the (parametrically large) $p$ fundamental strings that form it.~\footnote{See eq. (2.33) in~\cite{Martinec:2023plo}, and below.}
This is in harmony with the $R/\sqrt{\alpha'}\to\infty$ limit of~(\ref{e}),
\be\label{ade}
\lim_{{R\over\sqrt{\alpha'}}\to\infty}RE_w={\cal E}_{w,\rm string}~,
\ee
if and only if
\be\label{abtoone}
b(R/\sqrt{\alpha'}\to\infty)\to 1~,
\ee
or, equivalently,
\be\label{abzerotozero}
b_0(R/\sqrt{\alpha'}\to\infty)\to 0,
\ee
which we impose from now on.

For each set of $w=1,2,3,\dots$ long strings states with the same $j$ and other quantum numbers, we thus have, in particular,
\be\label{jj}
-{2j(j+1)\over k}+N_L+N_R-1={\cal E}_{\rm string}+{\rho_+^2+\rho_-^2\over 2\alpha'}~,
\ee
regardless of their winding number, $w$, where
\be\label{de1}
{\cal E}_{\rm string}\equiv{\cal E}_{1,\rm string}
\ee
is the energy carried by a string with $w=1$.

To recapitulate, we found that the long strings spectrum in superstring theory on deformed BTZ black holes has the pattern:
\be\label{recap}
E_w(R)+{wR\over\alpha'}(-1+b)=
-{wR\over\alpha'}+\sqrt{{w^2R^2\over\alpha'^2}+{2\over\alpha'}
\left({\cal E}_{\rm string}+{\rho_+^2+\rho_-^2\over 2\alpha'}\right)
+{n^2\over R^2}}~,
\ee
where ${\cal E}_{\rm string}$ is the dimensionless energy carried by a winding one long string in the superstring on a BTZ black-hole CFT:
\be\label{estring}
{\cal E}_{\rm string}=\lim_{R\to\infty}RE_1(R)~.
\ee
Note that the addition to ${\cal E}_{\rm string}$, in the parenthesis inside the square root on the r.h.s. of~(\ref{recap}),
is the dimensionless BTZ black-hole mass (in $1/\sqrt{\alpha' k}$ units) divided by $p$:~\footnote{See
eq. (2.23) in~\cite{Chakraborty:2023zdd}.}
\be\label{mbtza}
{\rho_+^2+\rho_-^2\over 2\alpha'}={r_5M_{BTZ}\over p}~.
\ee
All in all, this is the result in eq.~(\ref{patternn}).

In the negative case,~(\ref{nds})--(\ref{nphiphi}), repeating the above, gives rise to the on-shell condition
\be\label{nonshell}
{\alpha'\over 4}\left((E+bwR/\alpha')^2-p_{L,R}^2\right)-{j(j+1)\over k}+N_{L,R}={1\over 2}~,
\ee
instead of~(\ref{onshell}).~\footnote{More details:
the flip of signature of in the $(t,x)$ directions,~(\ref{nds}) versus~(\ref{ds}), amounts to
$$
g_{ij}=\pmatrix{
 1&0\cr
 0&-r_y^2&
}
$$
(instead of the one in the footnote below~(\ref{onshell})), and consequently to
$$
\Delta+\bar\Delta={1\over 2}\left((e+br_yw_y)^2-{n_y^2\over r_y^2}-w_y^2r_y^2\right)
$$
(instead of the the scaling dimension in that footnote),
in harmony with the first term in~(\ref{nonshell}).
}
From~(\ref{nonshell}), we get~\footnote{The choice of sign in front of the square root is for compatibility
with the negativity of the $b$ that is used in the negative coupling section,~(\ref{nbtoone}).}
\be\label{ne}
E_w+{wR\over\alpha'}(1+b)=
{wR\over\alpha'}-\sqrt{{w^2R^2\over\alpha'^2}-{2\over\alpha'}\left(-{2j(j+1)\over k}+N_L+N_R-1\right)+{n^2\over R^2}}~.
\ee
This is the result in eq.~(\ref{npattern}).

\end{document}